\documentclass[aps,prd,twocolumn,superscriptaddress,showpacs,showkeys]{revtex4-2}
\usepackage{color,xcolor}
\usepackage{amssymb}
\usepackage{graphicx}
\usepackage{extarrows}

\newcommand{\bastar}{\begin{eqnarray*}}
\newcommand{\eastar}{\end{eqnarray*}}
\newskip\humongous \humongous=0pt plus 1000pt minus 1000pt

\newif\ifdtup

\relax
\newcommand{\W}{{\vec W}}
\newcommand{\n}{\hat n}
\newcommand{\hn}{\hat n}
\newcommand{\hr}{\hat r}

\newcommand{\hD}{{\hat D}}

\newcommand{\bea}{\begin{eqnarray}}
\newcommand{\eea}{\end{eqnarray}}
\newcommand{\pd}{\partial}

\newcommand{\Int}{\displaystyle\int}
\newcommand{\A}{{\vec A}}
\newcommand{\hA}{{\hat A}}

\newcommand{\tA}{{\tilde A}}
\newcommand{\tC}{{\tilde C}}

\newcommand{\F}{{\vec F}}

\newcommand{\vPhi}{{\vec \Phi}}
\newcommand{\hF}{{\hat F}}

\newcommand{\mn}{{\mu\nu}}
\newcommand{\al}{\alpha}

\newcommand{\bae}{\bar e}

\newcommand{\e}{{\hat e}}

\newcommand{\bal}{{\bar \alpha}}

\newcommand{\vsig}{{\vec \sigma}}
\newcommand{\vphi}{{\varphi}}
\newcommand{\hth}{{\hat \theta}}
\newcommand{\hvp}{{\hat \varphi}}
\newcommand{\nn}{\nonumber}

\newcommand{\cL}{{\cal L}}

\newcommand{\cD}{{\cal D}}

\newcommand{\lam}{{\lambda}}

\begin{document}
\title{New Type of Monopoles in QCD and Standard Model}

\author{Liping Zou}
\email{Contact author: zoulp5@mail.sysu.edu.cn}
\affiliation{Sino-French Institute of Nuclear Engineering and Technology, 
Sun Yat-Sen University, Zhuhai 519082, China}
\author{Pengming Zhang}
\email{Contact author: zhangpm5@mail.sysu.edu.cn}
\affiliation{School of Physics and Astronomy, Sun Yat-Sen University, 
Zhuhai 519082, China}
\author{Y. M. Cho}
\email{Contact author: ymcho0416@gmail.com}
\affiliation{School of Physics and Astronomy, Seoul National University, Seoul 08826, Korea}
\affiliation{Korea Academy of Science and Technology, Seungnam 13630, Korea}

\begin{abstract}
We present the new monopole solutions in QCD, the Georgi-Glashow model, and the standard model. QCD has a new monopole, the Wu-Yang monopole which has the chromon dressing. The Georgi-Glashow model has a new monopole, the Wu-Yang monopole which has the W boson dressing, in addition to the well-known 'tHooft-Polyakov monopole. Moreover, the standard model has two types of monopoles, the well known Cho-Maison type monopoles which carry the magnetic charge $4\pi/e$ and the electromagnetically neutral type of magnetic monopoles (the naked one and the one with the W boson dressing) which carry the neutral magnetic charge $4\pi/\bae$, where $\bae$ is the neutral charge. We show how to construct the new monopole solutions and clarify the topological origin of the new monopoles. The new monopoles could have deep implications not just in high energy physics but also in condensed matter physics.
\end{abstract}
\keywords{Dirac monopole, Dirac charge quantization condition, Schwinger charge quantization condition, Wu-Yang monopole, 'tHooft-Polyakov monopole,  Cho-Maison monopole, Nambu monopole, Wu-Yang monopole with chromon dressing
in QCD, Wu-Yang monopole with W boson dressing in Georgi-Glashow model, neutral monopole in standard model, neutral charge quantization condition,  topology of the neutral monopole, mass of the neutral monopole}
\maketitle

\section{Introduction}

Since Dirac predicted the existence of 
the Dirac monopole generalizing the Maxwell's theory, the monopole has become an obsession in physics \cite{dirac,cab}. This is because this implies the existence of a totally new type of absolutely stable topological elementary particle in nature, which could explain the quantization of the electric charge and generate the hitherto unkown magnetic current. After the Dirac monopole, we have had the Wu-Yang monopole in QCD \cite{wu,prl80}, the 'tHooft-Polyakov monopole in Georgi-Glashow model \cite{tp}, the grand unification monopole in the grand unified theory \cite{dt},
and the Cho-Maison monopole in the standard model \cite{plb97,yang}. Among these the Cho-Maison monopole stands out as a most realistic monopole which could actually exist in nature, because this is predicted by the standard model. 

The standard model proposed by Weinberg has become a most successful theory in high energy physics \cite{wein}. The confirmation of the neutral current in the weak interaction by the Gargamelle experiment has made it the standard model of the electroweak interaction \cite{nc}. With the recent discovery of Higgs particle at LHC, it has been widely viewed that the theory has passed the ``final" test \cite{LHC}. This has urged people to go ``beyond" the standard model. 

Actually, however, the standard model has to pass another important test, the topological test. This is because it has a non-trivial monopole topology which predicts the existence of the Cho-Maison monopole. The importance of the Cho-Maison monopole is that it is the electroweak generalization of the Dirac monopole, so that it is this monopole (not the Dirac monopole) which could exist in nature. And this is within, not beyond, the standard model. This has not only triggered new studies on the electroweak monopole in the standard model \cite{chen,epjc15,
ellis,bb,mav,pta19,epjc20,gv,ms}, 
but also made MoEDAL and ATLAS at LHC and other experiments actively search for the monopole \cite{medal,atlas}. 
If detected, it will become the first absolutely stable magnetically charged topological elementary particle in the history of physics.  

The Cho-Maison monopole is well known to have the following properties \cite{plb97,pta19}. First, the magnetic charge is not $2\pi/e$ but $4\pi/e$, twice that of the Dirac monopole. So this is a Schwinger type monopole. This is because the period of the electromagnetic U(1) subgroup of the standard model becomes $4\pi$, not $2\pi$. Second, the mass of the monopole is estimated to be of the order of several TeV. This is because the mass basically comes from the same Higgs mechanism which makes the W boson massive, except that here the coupling is magnetic (i.e., $4\pi/e$). Third, in spite of this, the size of the monopole is set by the W boson mass. This is because the monopole solution has the weak boson dressing which sets the size by the W boson mass. Finally, it exists within (not beyond) the standard model as the electroweak generalization of the Dirac monopole. 

Because of these unique characteristics of the electroweak monopole MoEDAL and
ATLAS could detect the monopole without much difficulty, if LHC could produce it. However, the 14 TeV LHC may have no chance to produce the monopole if the mass becomes larger than 7 TeV. This is problematic, because in this case we may have to look for the remnant monopoles at present universe produced in the early universe to detect the monopole.

But the detection of the Cho-Maison monopole may not be the only 
topological test of the standard model. This is because the standard model might have another type of topological objects. To understand this, notice that the standard model has two conserved charges, the electric and neutral charges. In this case it could also have a neutral monopole which 
has the neutral magnetic charge $4\pi/\bae$, where $\bae$ is the neutral charge, in addition to the Cho-Maison monopole \cite{arx25}. This is a very interesting possibility which has never been discussed before. 

{\it The main purpose of this paper is to show that the standard model 
has two types of monopoles, the Cho-Maison type which carries the ordinary magnetic charge and the neutral type which carries the neutral magnetic charge. To show this we first show the existence of a new monopole 
in QCD, the Wu-Yang monopole which has the chromon dressing. Embedding this 
to the standard model, we obtain 
the neutral monopole in the standard model. Moreover, we argue that the mass of the neutral monopole could be as low as 3.28 TeV, lower than the mass of the Cho-Maison monopole. With the neutral monopole we have two charge quantization conditions from the standard model, the neutral charge quantization condition $\bar e =(4\pi/\bar g) n$ as well as the Dirac charge quantization condition  $e =(4\pi/g) n$. We show how to construct such monopoles and discuss the topology of the new monopoles. Our result could change the experimental detection of the electroweak monopole at LHC drastically.} 

In doing so, we show a new monopole in Georgi-Glashow model, the Wu-Yang monopole which has only the W boson dressing. We also show that the 'tHooft-Polyakov monopole is nothing but the Wu-Yang monopole dressed by 
the Higgs and W bosons which make the monopole energy finite. Moreover, in the standard model we show the existence of three new monopoles,
two neutral type (the naked one and the one dressed by the W boson) and an extra Cho-Maison type which has only the W boson dressing.   

The existence of the new monopoles in QCD and Georgi-Glashow model is very  interesting from the theoretical
point of view. But the new monopoles 
in the standard model could play more important roles because they are the realistic monopoles which could actually exist in nature and be detected. The fact that the standard model actually accommodates two types of monopoles is totally unexpected. Moreover, the neutral monopoles are a totally new species of monopoles whose existence has never been discussed in physics before. 

In this respect it becomes important 
to estimate the mass of the neutral monopole in the standard model. A simple and quick way to estimate this mass is to adopt the argument proposed by Zeldovich and Khlopov \cite{zel}. In this case one might predict the mass of the neutral monopole to be near $1/\bal$ times the W boson mass, around 3.28 TeV, where $\bal=\bae^2/4\pi$ is the fine structure constant of the neutral interaction. This is interesting, because this suggests that the neutral monopole mass could be smaller than the Cho-Maison monopole mass. 

This could change the topological test of the standard model completely. This is because now we (not only have more topological objects but also) have the possibility that the neutral monopole mass could be smaller than the Cho-Maison monopole mass. If this is true, the present 14 TeV LHC could produce them in pairs, and in principle ATLAS and MoEDAL could detect this. This implies that the first topological test of the standard model could come from the detection of the neutral monopole, not the Cho-Maison monopole. 

The existence of the new monopoles in 
the standard model may also play important roles in low energy physics. This is because the standard model could also have important applications in condensed matter physics. For example, it could be interpreted as 
the non-Abelian Ginzburg-Landau theory of two-gap ferromagnetic superconductivity \cite{pla23,ap24,
arx251}. This is because the Higgs doublet could be viewed to describe 
the spin doublet Cooper pair in two 
gap ferromagnetic superconductors. In this view the hypercharge U(1) gauge interaction describes the electromagnetic interaction acting on the charged spin doublet Cooper pair, and the weak SU(2) gauge interaction describes the non-Abelian magnon interaction acting on the spin of the doublet Cooper pair.

Moreover, the standard model could 
also be interpreted as a non-Abelian gauge theory of electron spintronics \cite{arx251,arx252,arx253}. This is because the Higgs doublet could also 
be interpreted as the charged spinon 
in electron spintronics. So, with the hypercharge U(1) interaction as the electromagnetic interaction acting on the electron and the weak SU(2) interaction as the non-Abelian magnon gauge interaction acting on 
the electron spin, one could interpret 
it as a theory of the electron magnon 
spintronics. This strongly implies that the new topological objects in the standard model could also play important roles in the low energy physics.

The paper is organized as follows. In Section II we review the Abelian decomposition of SU(2) QCD and show 
the existence of the new monopole in QCD, the Wu-Yang monopole which has 
the chromon dressing. In Section III 
we discuss the new monopole in Georgi-Glashow model, the Wu-Yang monopole which has the W boson dressing. In Section IV we review the Abelian decomposition of the standard model to clarify the hidden structure of the theory. In Section V we show that 
the standard model has another new Cho-Maison monopole, the Cho-Maison monopole which has only the W boson dressing. In Section VI we show that 
the standard model has a totally new type of monopoles, the neutral 
monopoles which carry the neutral magnetic charge, and show why they should exist. Finally in the last section we discuss the physical implications of the new monopoles.  

\section{New Monopole in QCD: Wu-Yang Monopole with Chromon Dressing} 

It has been well known that QCD has 
the Wu-Yang monopole \cite{wu,prl80}. This monopole plays an important role, because this could make the monopole condensation and generate the dual Meissner effect needed for the color confinement in QCD \cite{prd13,epjc19}. In the following we show that QCD 
has a new monopole, the generalized 
Wu-Yang monopole which has the chromon dressing. To do this we need the Abelian decomposition of the SU(2) 
QCD first.

Let $(\n_1,\hn_2,\hn_3)$ be an arbitrary orthonormal basis of SU(2). 
Selecting $\n=\n_3$ to be the Abelian direction, we have the Abelian decomposition of the color gauge potential $\A_\mu$ to the restricted part $\hA_\mu$ and the valence part $\W_\mu$ \cite{prd80,prl81,fadd,shab,
zucc,kondo}, 
\begin{gather}
\A_\mu = \hA_\mu + \W_\mu,     \nn\\
\hA_\mu = A_\mu \n -\frac1g \hn
\times \pd_\mu \n
= \tA_\mu +\tC_\mu, \nn\\
\tA_\mu=A_\mu \n,~~~\tC_\mu=-\frac{1}{g} \n\times \pd_\mu \n,
~~~(A_\mu = \n \cdot \A_\mu),  \nn\\
\W_\mu =W^1_\mu~\n_1 + W^2_\mu~\n_2,
~~~(\n\cdot \W_\mu =0).
\label{adec}
\end{gather}
This is the Abelian decomposition of  QCD which decomposes the gluon potential to the non-topological Maxwellian Abelian potential $\tA_\mu$ which describes the color neutral nuron, the gauge covariant valence potential $\W_\mu$ which describes 
the colored chromon, and the topological Diracian potential $\tC_\mu$ which describes the 
non-Abelian monopole. In this decomposition the gluons are decomposed to the color neutral nuron and colored chromon gauge independently.

With this Abelian decomposition 
we have
\begin{gather}
\F_\mn=\hF_\mn + \hD _\mu \W_\nu - \hD_\nu
\W_\mu + g\W_\mu \times \W_\nu,   \nn\\
\hD_\mu=\pd_\mu+g \hA_\mu \times, \nn\\
\hF_\mn= \pd_\mu \hA_\nu-\pd_\nu \hA_\mu
+ g \hA_\mu \times \hA_\nu =F_\mn' \n, \nn \\
F'_\mn=F_\mn + H_\mn
= \pd_\mu A'_\nu-\pd_\nu A'_\mu,  \nn\\
F_\mn =\pd_\mu A_\nu-\pd_\nu A_\mu,  \nn\\
H_\mn = -\frac1g \n \cdot (\pd_\mu \n \times \pd_\nu \n)
=\pd_\mu C_\nu-\pd_\nu C_\mu,  \nn\\
A_\mu' = A_\mu+ C_\mu,
~~~C_\mu \simeq -\frac1g \n_1 \cdot \pd_\mu \n_2.
\label{cdec}
\end{gather}
Notice that $C_\mu$ is determined up to 
the U(1) gauge transformation which 
leaves $\n$ invariant. 

From this we can construct the restricted QCD (RCD) which describes the Abelian subdynamics of QCD which nevertheless has the full non-Abelian color gauge invariance,
\begin{gather}
\cL_{RCD} =-\frac14 \hF_\mn^2
=-\frac14 (F_\mn +H_\mn)^2. 
\label{rcd}	
\end{gather}
Moreover we	can express the QCD Lagrangian by
\begin{gather} 
\cL_{QCD} = -\frac14 \F^2_\mn \nn\\
=\cL_{RCD} -\frac14 (\hD_\mu\W_\nu-\hD_\nu\W_\mu)^2 \nn\\
-\frac{g}{2} {\hF}_\mn \cdot (\W_\mu \times \W_\nu)
-\frac{g^2}{4} (\W_\mu \times \W_\nu)^2  \nn\\
=-\frac14 (F_\mn +H_\mn + W_\mn)^2
-\frac14 (\hD_\mu\W_\nu
-\hD_\nu\W_\mu)^2,    \nn\\
W_\mn= g \n \cdot (\W_\mu \times \W_\nu). 
\label{ecd} 
\end{gather}
This tells that QCD can be viewed as RCD made of the restricted potential $\hA_\mu$ which has the gauge covariant chromon $\W_\mu$ as the colored source. 

The Lagrangian (\ref{ecd}) gives us the following equation of motion for SU(2) QCD,
\begin{gather}
\pd_\mu (F_\mn +H_\mn)  \nn\\
= -g\n \cdot \big[\W_\mu \times (\hD_\mu\W_\nu-\hD_\nu\W_\mu) 
+\pd_\mu (\W_\mu \times \W_\nu) \big], \nn\\
\hD_\mu (\hD_\mu\W_\nu-\hD_\nu\W_\mu) \nn\\
= g (F_\mn +H_\mn +W_\mn)~\n \times \W_\mu.
\label{ecdeq}
\end{gather}
This should be compared with the well-known QCD equation of motion 
\begin{gather}
D_\mu \F_\mn =0.
\label{qcdeq}
\end{gather} 
So the Abelian decomposition decomposes the QCD equation (\ref{qcdeq}) to two separate equations, the one for the nuron $A_\mu$ and the one for the chromon $\W_\mu$. 

Moreover, the nuron equation is very 
much like the Maxwell's equation in 
QED, which tells that the nuron plays 
the role of ``the photon" or the binding gluon in QCD. And the chromon equation tells that the chromon becomes the colored source of the nuron, which plays the role of the constituent gluon in 
QCD \cite{prd80,prl81}. 

Most importantly this tells that QCD has the conserved current made of the chromon which assures the existence of the conserved color charge. This is because the first equation of (\ref{ecdeq}) can be expressed by
\begin{gather}
\pd_\mu (F_\mn +H_\mn) =J_\nu^{qcd}, \nn\\
J_\mu^{qcd} =-g\n \cdot \big[\W_\nu \times (\hD_\nu\W_\mu-\hD_\mu\W_\nu) \nn\\
+\pd_\nu (\W_\nu \times \W_\mu) \big]. 
\label{ecdj}
\end{gather}
This tells that the Abelian decomposition decomposes QCD to the Abelian part and valence part in such a way that the color conservation is made explicit. Notice that the existence of the conserved 
color charge in QCD is completely obsecured in (\ref{qcdeq}).

To obtain a new monopole solution, we choose the ansatz
\begin{gather}
\hA_\mu =-\frac1g \hr \times\pd_\mu \hr,
~~~\W_\mu =\frac{1}{g} f(r)\hr \times\pd_\mu \hr,
\label{cmans}
\end{gather}
and have the following equation from (\ref{ecdeq}) 
\begin{gather}
\ddot f -\frac{f^2 -1}{r^2}~f =0.
\label{qcdmeq}
\end{gather} 
This has three trivial solutions 
\begin{gather}
f=+1,~~~f=0,~~~f=-1.
\label{wy}
\end{gather} 
The first one is the trivial vacuum
$\A_\mu=0$, and the thrid one describes another vacuum
\begin{gather}
\A_\mu =-\frac{2}{g} \hr \times\pd_\mu \hr.
\label{vac}
\end{gather}
This is because in the matrix notation this is expressed by
\begin{gather}
\bar A_\mu =-\frac{2}{g}  
(\hr \times\pd_\mu \hr)\cdot \frac{\vsig}{2}
=\frac{i}{g} U \pd_\mu U^{-1},  \nn\\
U= i\hr \cdot \vsig,
~~~U^\dag =-i\hr \cdot \vsig =U^{-1}.
\end{gather}
So both $f=+1$ and $f=-1$ describe 
a vacuum.

\begin{figure}
\includegraphics[height=4.5cm, width=8cm]{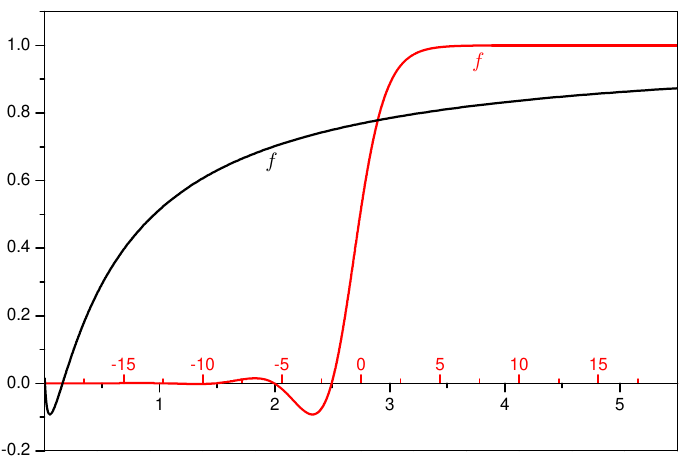}
\caption{\label{nqcdm} The new monopole solution in QCD, the Wu-Yang monopole which has the chromon dressing. The black curve represents the chromon profile, and the red curve represents the chromon profile $f$ in the log scale.}
\end{figure}

The second one, of course, is the Wu-Yang monopole in QCD \cite{prd80,
prl81,prl80,wu},
\begin{gather}
\hA_\mu =-\frac{1}{g} \hr \times\pd_\mu \hr,~~~\W_\mu =0.
\label{wy}
\end{gather}
We can easily show that this Wu-Yang monopole is nothing but the Dirac monopole embedded in QCD. To show this, notice that under the gauge transformation
\begin{gather}
\hr =\left(\begin{array}{ccc}
\sin \theta \cos \vphi \\ 
\sin \theta \sin \vphi \\ 
\cos \theta \end{array} \right)
\rightarrow \e =U \hr,  
\label{gt0}
\end{gather} 
we have
\begin{gather}
\hA_\mu =-\frac{1}{g}\hr \times\pd_\mu \hr \rightarrow C_\mu~\e, 
~~~\W_\mu  \rightarrow 0,  \nn\\
C_\mu \simeq-\frac1g \hth \cdot \pd_\mu \hvp
=-\frac1g (1-\cos \theta)\pd_\mu \vphi,  \nn\\
\hth =\left(\begin{array}{ccc}
\cos \theta \cos \vphi \\
\cos \theta \sin \vphi \\
-\sin \theta \end{array} \right),
~~~\hvp =\left(\begin{array}{ccc}
-\sin \vphi \\ \cos \vphi \\
0 \end{array} \right).
\label{dpot}
\end{gather}
This is the Dirac monopole trivially embedded in QCD, which confirms that the Wu-Yang monopole is nothing but the Dirac monopole embedded in the SU(2) QCD.

Now, we can generalize this Wu-Yang monopole solution with non-vanishing $f$. Indeed, solving (\ref{qcdmeq}) 
with the boundary condition
\begin{gather}	
f(0) = 0,~~~f(\infty) =1,
\label{qcdmbc}
\end{gather}
we obtain a new monopole solution in QCD, the Wu-Yang monopole which has the chromon dressing. This is shown in Fig. \ref{nqcdm}.

Near the origin the solution has the following behavior,
\begin{gather}
f(r)\simeq r^{1/2}\Big(A\cos\big(\frac{\sqrt{3}}{2}\ln r \big) +B\sin\big(\frac{\sqrt{3}}{2}\ln r \big) \Big)\nn\\
+O\left(r^{3/2}\right),
\end{gather}
where $A$ and $B$ are the integration constants. This explains the interesting behavior of the chromon profile near 
the origin shown in red curve in 
Fig. \ref{nqcdm}. But asymptotically 
it behaves as
\begin{gather}
f(r)\simeq 1
-\frac{C}{r}+O\Big(\frac{1}{r^2} \Big),
\end{gather}
where $C$ is another integration constant. This tells that the monopole solution approaches to 
the trivial vacuum at the infinity, but it has $1/r$ tail and has 
no fixed size. 

We can calculate the magnetic charge of the dressed monopole. Since the magnetic flux of the monopole is given by the topological Diracian part $H_\mn$ of the gauge independent Abelian field strength $F_\mn'$ of 
the solution, 
\begin{gather}
H_\mn= -\frac1g \sin \theta 
\big(\pd_\mu \theta \pd_\nu \vphi
-\pd_\nu \theta \pd_\mu \vphi \big),  
\label{qcdmf}
\end{gather}
we have   
\begin{gather}	
q_m = \int_{r=\infty}
H_{\theta \vphi} d\theta d \vphi 
= -\frac{4\pi}{g}.  
\label{qcdmc}
\end{gather}	
This, of course, is exactly the magnetic charge of the Wu-Yang monopole. This
tells that the chromon dressing does 
not change the monopole charge.  

The energy of the monopole is given by
\begin{gather}
E = E_{WY} + E_C,   \nn\\
E_{WY} = \frac{4\pi}{g^2}
\Int_0^\infty~\frac{dr}{2r^2}, \nn\\	
E_C =\frac{4\pi}{g^2} \Int_0^\infty 
\Big[\dot f^2 
+\frac{f^2-2}{2r^2} f^2 \Big] dr,
\label{qcdme}
\end{gather}
where $E_{WY}$ is the energy of Wu-Yang monopole and $E_C$ is the energy of the chromon dressing. Both are divergent near the origin. But both are convergent and has the same behavior near the infinity, which assures that 
the solution is physical.

The importance of the chromonic monopole in QCD is that it is a prototype non-Abelian monopole which appears in almost all SU(2) gauge theories, including the Georgi-Glashow model and the standard model, as we will see in the following. Of course the monopole has different physical interpretations in different models, but mathematically 
exactly the same chromonic monopole of QCD appears as the monopole solution in these non-Abelian gauge theories.

\section{New Monopole in Georgi-Glashow Model}

It is well known that the Georgi-Glashow model has the 'tHooft-Polyakov monopole. Now we show that this model has a new monopole solution different from the 'tHooft-Polyakov monopole. Let us start from the Georgi-Glashow Lagrangian
\begin{gather}
\cL_{GG} =-\frac12 (D_\mu \vPhi)^2
-\frac{\lam}{4}  \big(\vPhi^2
-\frac{\mu^2}{\lam} \big)^2  
-\frac14 \F_\mn^2,
\label{ggl1}
\end{gather}
where $\vPhi$ is the Higgs triplet. 
With
\begin{gather}
\vPhi =\rho \n,
~~~~\A_\mu=\hA_\mu +\W_\mu,
\end{gather}
we have the Abelian decomposition of the Georgi-Glashow Lagrangian
\begin{gather}
\cL_{GG} =-\frac12 (\pd_\mu \rho)^2
-\frac{\lam}{4} \big(\rho^2 
-\frac{\mu^2}{\lam} \big)^2
-\frac14 \hF_\mn^2  \nn\\
-\frac14(\hD_\mu \W_\nu -\hD_\nu \W_\mu)^2 -\frac{g^2}{2} {\rho}^2 (\W_\mu)^2   \nn\\
-\frac{g}{2} \hF_\mn \cdot(\W_\mu \times \W_\nu)
-\frac{g^2}{4} (\W_\mu \times \W_\nu)^2  \nn\\ 
= -\frac12 (\pd_\mu \rho)^2
-\frac{\lam}{4}\big(\rho^2 
-\bar \rho_0^2 \big)^2 -\frac14 {F'}_\mn^2 \nn\\
-\frac12 |D'_\mu W_\nu -D'_\nu W_\mu|^2
-g^2{\rho}^2 W_\mu^* W_\mu 
+ ig {F'}_\mn W_\mu^*W_\nu  \nn\\
+ \frac{g^2}{4}(W_\mu^*W_\nu 
-W_\nu^* W_\mu)^2,
\label{ggl2}
\end{gather}
where 
\begin{gather}
\bar \rho_0 =\sqrt{\mu^2/\lam},
~~~W_\mu =\frac{1}{\sqrt 2} (W^1_\mu 
+i W^2_\mu),      \nn\\
\hD_\mu = \pd_\mu + g \hA_\mu \times, 
~~~~D'_\mu = \pd_\mu +ig A'_\mu.
\end{gather}
This shows that the theory can be viewed as an Abelian gauge theory 
which has the charged vector field $W_\mu$ as a source. 

In this Abelianization the Higgs triplet $\vphi$ disappears completely. 
In stead we hace the scalar Higgs field $\rho$ which has the mass $M_H =\sqrt{2\lam}\bar\rho_0=\sqrt{2}\mu$.
Moreover, the gauge boson $\W_\mu$
acquires the mass $M_W =g \bar \rho_0$,
absorbing the two degrees of the Higgs triplet. This, of course, is the Higgs mechanism. But here we do not need any symmetry breaking, spontaneous or not, to have the mass generation. To see 
this notice that (\ref{ggl2}) is mathematically identical to (\ref{ggl1}), so that it retains 
the full non-Abelian $SU(2)$ gauge symmetry of (\ref{ggl1}). And here 
the W boson acquire mass from the non-vanishing vacuum value of 
the Higgs field $\rho_0$, which does not break any symmetry. In fact, in (\ref{ggl2}) we have no Higgs doublet which can break the $SU(2)$ symmetry 
at all. This confirms that we do 
not need any spontaneous symmetry 
breaking to have the mass 
generation \cite{epjc15,pta19,epjc20}.  

From the Lagrangian (\ref{ggl2}) we have the equation of motion
\begin{gather}
\pd^2\rho -2g^2 \rho~W_\mu^* W_\mu 
=\lam \big(\rho^2-\rho_0^2 \big) \rho, \nn\\
D_\mu'(D_\mu'W_\nu-D_\nu'W_\mu)
=g^2 \rho^2 W_\nu +ig F'_\mn W_\mu \nn\\
-g^2 (W_\mu^* W_\nu -W_\nu^* W_\mu) W_\mu,   \nn\\
\pd_\mu F'_\mn =ig \pd_\mu (W_\mu^*W_\nu
-W_\nu^* W_\mu)
+ig \big[W_\mu^* (D'_\mu W_\nu \nn\\
-D'_\nu W_\mu) +(D'_\mu W_\nu
-D'_\nu W_\mu)^* W_\mu \big].
\label{ggeom2}
\end{gather}
Notice that here $A_\mu'$ and $W_\mu$ have two independent equations of motion.

To discuss the 'tHooft-Polyakov monopole we choose the ansatz, 
\begin{gather}
	\vPhi=\rho(r)~\hr, 
	~~~\hA_\mu=-\frac{1}{g}\hr \times \pd_\mu \hr, \nn\\
	\W_\mu =\frac1g f(r)~\hr \times \pd_\mu \hr,
	\label{tpans0}
\end{gather}
or equivalently
\begin{gather}
	\rho = \rho(r), 
	~~~A'_\mu =-\frac{1}{g}(1-\cos\theta) 
	\pd_\mu \varphi,  \nn\\
	W_\mu =\frac{i}{g}\frac{f(r)}{\sqrt2}
	(\pd_\mu \theta +i \sin\theta \pd_\mu \vphi).
	\label{tpans1}
\end{gather}
This should be compared to the monopole ansatz (\ref{cmans}) of QCD. Clearly this becomes identical to (\ref{cmans}) when $\rho$ becomes zero.

With the ansatz we have the following equation for the monopole 
\begin{gather}
\ddot{\rho}+\frac{2}{r}\dot{\rho} 
- 2\frac{f^2}{r^2}\rho 
=\lam \big(\rho^2 -\bar \rho_0^2 \big) \rho, \nn\\
\ddot{f} -\frac{f^2-1}{r^2}f 
=g^2\rho^2 f.
\label{tpmeq}
\end{gather}
This has the solution
\begin{gather}
\rho =\bar \rho_0,~~~~f=0, 
\label{wym}
\end{gather}
or
\begin{gather}
\rho =\bar \rho_0,
~~~\hA_\mu =-\frac{1}{g}\hr \times\pd_\mu \hr, 
~~~\W_\mu =0.
\label{wym}
\end{gather}
This of course is the Wu-Yang monopole \cite{prd80,prl81,prl80,wu}. But this has an infinite energy.

\begin{figure}
\includegraphics[height=4.5cm, width=8cm]{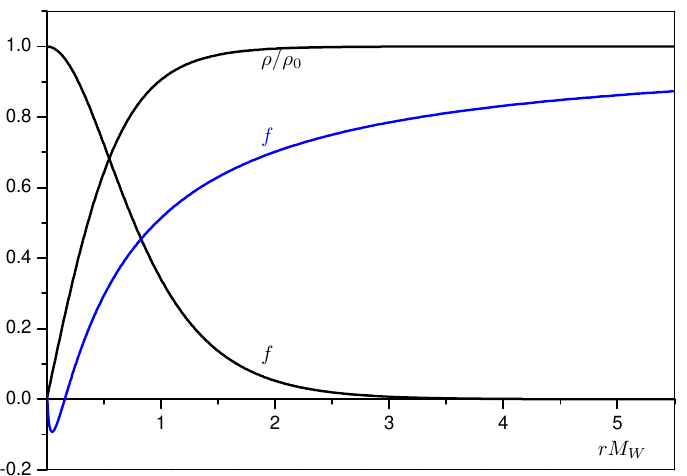}
\caption{\label{ntpm} The new monopole solutions in Georgi-Glashow model. 
The black curves represent the well 
known 'tHooft-Polyakov monopole, and 
the blue curve represents the Wu-Yang monopole which has only the W boson dressing.}
\end{figure}

We can make the energy finite with 
the W boson dressing. Integrating (\ref{tpmeq}) with the boundary condition
\begin{gather}
	\rho(0)=0,~~~~\rho(\infty)=\bar \rho_0, \nn\\
	f(0)=1,~~~~f(\infty)=0,
	\label{ggbc}
\end{gather}
we obtain the 'tHooft-Polyakov monopole which has a finite energy. The solution is shown in Fig. \ref{ntpm} in black curves. 

The energy of the 'tHooft-Polyakov monopole is given by
\begin{gather}		
E =E_\rho +E_W +E_{\rho W},  \nn\\	
E_\rho = M \Int_0^\infty \frac{r^2}{2} 
\Big[(\frac{\dot\rho}{\bar \rho_0})^2
+\frac{\lam}{2}
\big((\frac{\rho}{\bar \rho_0})^2-1 \big)^2 \Big] M_W dr, \nn\\
E_W =M \Int_0^\infty \Big[\dot f^2 +\frac{(f^2-1)^2}{2r^2} \Big] 
\frac{dr}{M_W}, \nn\\
E_{\rho W} =M \Int_0^\infty f^2 (\frac{\rho}{\bar \rho_0})^2 M_W dr,  \nn\\
M =\frac{4\pi}{g^2} M_W.
\label{tpme}
\end{gather}		
Notice that the energy scale is set by $M$. The boundary condition $f(0)=1$ is crucial to make the solutions regular at the origin and make the energy finite. This tells that the 'tHooft-Polyakov monopole is nothing but the Wu-Yang monopole which has the Higgs and W boson dressing which makes the energy finite.

Moreover, in the Bogomol'nyi-Prasad-Sommerfield 
(BPS) limit when $\lam$ goes to zero, the equation (\ref{tpmeq}) reduces to the following equation 
\begin{gather}
\dot{\rho}\pm \frac{f^2-1}{gr^2}=0,
~~~\dot{f} \pm g \rho f=0. 
\label{pseq}
\end{gather}
This has the analytic solution
\begin{gather}
\rho= \rho_0\coth(g \bar \rho_0 r)-\dfrac{1}{gr},
~~~f= \dfrac{g \bar \rho_0 r}{\sinh(g \bar \rho_0 r)},
\end{gather}
which describes the analytic Prasad-Sommerfield monopole \cite{ps}. And we can generalize 
the 'tHooft-Polyakov monopole to the dyon and anti-dyon. 

Now, we show that the Georgi-Glashow model (just like QCD) has another interesting new monopole solution. To see this, notice that (\ref{tpmeq}), in the absence of the Higgs triplet $\vPhi$ (i.e., with $\rho=0$) the monopole ansatz (\ref{tpans1}) reduces to 
\begin{gather}
\rho = 0, 
~~~A'_\mu =-\frac{1}{g}(1-\cos\theta) 
\pd_\mu \varphi,  \nn\\
W_\mu =\frac{i}{g}\frac{f(r)}{\sqrt2}
(\pd_\mu \theta +i \sin\theta \pd_\mu \vphi).
\label{ntpans}
\end{gather}
Moreover, the monopole equation 
(\ref{tpmeq}) reduces to 
\begin{gather}
\ddot f -\frac{f^2 -1}{r^2}~f =0.
\label{ntpmeq}
\end{gather}
This is identical to the equation (\ref{qcdmeq}) we have discussed above in QCD. 

Integrating this with the boundary condition (\ref{qcdmbc}) we obtain a new monopole solution, the Wu-Yang monopole which has only the W boson dressing. This is shown in Fig. \ref{ntpm} in blue curves. This tells that the Georgi-Glashow model also admits a new type of monopole solution which is mathematically identical to the chromonic monopole in QCD.

The energy of the new monopole is given by
\begin{gather}	
E = E_{WY} + E_W,   \nn\\
E_{WY} = \frac{4\pi}{g^2}
\Int_0^\infty~\frac{dr}{2r^2},  \nn\\	
E_W =\frac{4\pi}{g^2} \Int_0^\infty 
\Big[\dot f^2 
+\frac{f^2-2}{2r^2} f^2 \Big] dr,  
\end{gather}		
where $E_{WY}$ is the energy of the Wu-Yang monopole and $E_W$ is the energy of the W boson dressing. Both are divergent, so that the new monopole in Georgi-Glashow model also has an infinite energy. The existence of the new monopole in  Georgi-Glashow model, the Wu-Yang monopole dressed by the W boson, is also completely unexpected. 

\section{New Face of Standard Model: Two Conserved Currents and Their Mixing}

The new monopole solutions in QCD and Georgi-Glashow model are not the only new monopoles in non-Abelian gauge theories. Remarkably, the standard model also has the new type of monopole solution, the neutral magnetic monopole which carries the neutral magnetic charge. This means that the standard model actually has two types of monopoles, the Cho-Maison type and 
the neutral type. This is really unexpected. 

To discuss the new monopoles, we have to understand the hidden structure of the standard model first. For this we need the Abelian decomposition of the standard model. Consider the (bosonic sector of) Weinberg-Salam model,
\begin{gather}
{\cal L} =-|{\cal D}_\mu \phi|^2
-\frac{\lambda}{2}\big(|\phi|^2
-\frac{\mu^2}{\lambda}\big)^2
-\frac14\F_\mn^2
-\frac{1}{4}G_\mn^2, \nn \\
{\cal D}_\mu \phi =\big(\pd_\mu
-i\frac{g}{2} \vsig \cdot \A_\mu
-i\frac{g'}{2} B_\mu\big) \phi  \nn\\
=D_\mu \phi-i\frac{g'}{2} B_\mu \phi,
\label{lag0}
\end{gather}
where $\phi$ is the Higgs doublet, $\A_\mu$, $\F_\mn$ and $B_\mu$, $G_\mn$ are the gauge fields of $SU(2)$ and hypercharge U(1), and $D_\mu$ is the covariant derivative of 
the weak SU(2). Expressing $\phi$ by 
the scalar Higgs boson $\rho$ and 
unit doublet $\xi$ by
\begin{gather}
\phi = \dfrac{1}{\sqrt{2}} \rho~\xi,~~~(\xi^\dagger \xi = 1),
\end{gather}
we have
\begin{gather}
{\cal L}=-\frac{1}{2} (\pd_\mu \rho)^2
- \frac{\rho^2}{2} |{\cal D}_\mu \xi |^2
-\frac{\lambda}{8}\big(\rho^2-\rho_0^2 \big)^2 \nn\\
-\frac14 \F_\mn^2 -\frac14 G_\mn^2,
\label{lag1}
\end{gather}
where $\rho_0=\sqrt{2\mu^2/\lambda}$ is 
the vacuum value of the Higgs field. Notice 
that the hypercharge U(1) coupling of $\xi$ makes the theory a gauge theory of $CP^1$ field \cite{plb97}.

Choosing an Abelian direction $\n$, 
we have the Abelian decomposition of the SU(2) gauge potential $\A_\mu$ to 
the restricted part $\hA_\mu$ and the valence part $\W_\mu$ as we did in (\ref{adec}) and (\ref{cdec}) in QCD.  
Moreover, since $\n$ is arbitrary, we can choose $\n=-\xi^\dag \vsig \xi$. With this we have
\begin{gather}
C_\mu \simeq -\frac{2i}{g} \xi^\dagger \pd_\mu \xi,  \nn\\	
\cD_\mu \xi= \big[-i\frac{g}{2}(A'_\mu \n
+\W_\mu) \cdot \vsig 
-i\frac{g'}{2 } B_\mu \big]~\xi, \nn\\
|\cD_\mu \xi|^2 =\frac{1}{4} (-gA'_\mu+g'B_\mu)^2
+\frac{g^2}{4} \W_\mu^2.
\label{id}
\end{gather}
Using this we can remove the unit doublet $\xi$ from the Weinberg Lagrangian and ``abelianize" it,  
\begin{gather}
\cL = -\frac12 (\pd_\mu \rho)^2
-\frac{\lam}{8}\big(\rho^2-\rho_0^2 \big)^2 \nn\\
-\frac14 {F_\mn'}^2 -\frac14 {G_\mn}^2
-\frac12 \big|D_\mu' W_\nu 
-D_\nu' W_\mu \big|^2  \nn\\	
-\frac{\rho^2}{8} \big[(-gA_\mu'+g'B_\mu)^2 
+2 g^2 W_\mu^*W_\mu \big]  \nn\\
+i g F_\mn' W_\mu^* W_\nu 
+ \frac{g^2}{4}(W_\mu^* W_\nu 
-W_\nu^* W_\mu)^2,  \nn\\
D_\mu' =\pd_\mu +ig A_\mu',
~~~W_\mu =\frac{1}{\sqrt 2} (W^1_\mu 
+i W^2_\mu).
\label{lag2}
\end{gather}
Notice that when we switch off the hypercharge gauge interaction, the Weinberg Lagrangian reduces to
\begin{gather}
\cL = -\frac12 (\pd_\mu \rho)^2
-\frac{\lam}{8}\big(\rho^2-\rho_0^2 \big)^2 -\frac14 {F_\mn'}^2 
-\frac{g^2}{8} \rho^2 A_\mu'^2 \nn\\
-\frac12 \big|D'_\mu W_\nu 
-D'_\nu W_\mu \big|^2  
-\frac{g^2}{4}\rho^2 W_\mu^*W_\mu  \nn\\
+i g F_\mn' W_\mu^* W_\nu 
+ \frac{g^2}{4}\big(W_\mu^* W_\nu 
-W_\nu^* W_\mu \big)^2.
\label{lag4}
\end{gather}  
This is almost identical to the  Georgi-Glashow Lagrangian (\ref{ggl2}), 
except that in (\ref{lag4}) the Abelian potential $A_\mu'$ of SU(2) has the extra interaction with the Higgs scalar which generates the mass to $A'_\mu$.

From this we have the following equation of motion,
\begin{gather}
\pd_\mu^2\rho-\frac{\rho}{4}\Big((gA'_\mu
-g'B_\mu)^2+2g^2W_\mu^* W_\mu \Big)
=\frac{\lam}{2}(\rho^2-\rho_0^2) \rho, \nn\\
D_\mu'\Big(D_\mu'W_\nu -D_\nu'W_\mu \Big)
=\frac{g^2}{4}\rho^2 W_\nu  \nn\\
+igF_\mn' W_\mu +g^2 \Big(W_\mu^* W_\nu
-W_\nu^* W_\mu\Big) W_\mu,  \nn\\
\pd_\mu F_\mn' = j_\nu^{(w)},  \nn\\
\pd_\mu G_\mn = j_\nu^{(h)},
\label{wseq1}
\end{gather}
where
\begin{gather}
j_\nu^{(w)} =ig\Big[\pd_\mu(W_\mu^* W_\nu
-W_\nu^* W_\mu) + W_\mu^*(D_\mu'W_\nu 
-D_\nu'W_\mu) \nn\\
-W_\mu (D_\mu'W_\nu -D_\nu'W_\mu)^*  -\frac{i}{4}(gA_\nu'-g'B_\nu) \rho^2 \Big],  \nn\\
j_\nu^{(h)} =-\frac{g'}{4}(gA_\nu'
-g'B_\nu) \rho^2.
\label{j1}
\end{gather}
Notice that the last two equations of
(\ref{wseq1}) tells that the theory has two conserved currents, the weak current $j_\mu^{(w)}$ of $A_\mu'$ and the hypercharge current $j_\mu^{(h)}$ of $B_\mu$. We emphasize that the existence of two conserved currents (\ref{j1})
is completely obsecured in the Weinberg
Lagrangian (\ref{lag0}). Only after 
the Abelian decomposition the existence of two conserved currents is made explicit.  

The Lagrangian (\ref{lag2}) tells that $A_\mu'$ and $B_\mu$ are not the mass eigenstates. To express Lagrangian 
in terms of physical fields we introduce the Weinberg mixing and define $A_\mu^{\rm (em)}$ and $Z_\mu$ 
by
\begin{gather}
\left(\begin{array}{cc} A_\mu^{\rm (em)} \\  Z_{\mu}
\end{array} \right)
=\frac{1}{\sqrt{g^2 + g'^2}} \left(\begin{array}{cc} g & g' \\
-g' & g \end{array} \right)
\left(\begin{array}{cc} B_{\mu} \\ A'_{\mu}
\end{array} \right)  \nn\\
= \left(\begin{array}{cc}
\cos\theta_{\rm w} & \sin\theta_{\rm w} \\
-\sin\theta_{\rm w} & \cos\theta_{\rm w}
\end{array} \right)
\left( \begin{array}{cc} B_{\mu} \\ A_\mu'
\end{array} \right).
\label{mix}
\end{gather}
Now, with the identity (\ref{id}) we can 
express (\ref{lag2}) by
\begin{gather}
\cL= -\frac12 (\pd_\mu \rho)^2
-\frac{\lam}{8}\big(\rho^2-\rho_0^2 \big)^2
-\frac14 {F_\mn^{\rm (em)}}^2   \nn\\
-\frac12 \big|(D_\mu^{\rm (em)} 
+i\bae Z_\mu) W_\nu
-(D_\nu^{\rm (em)} +i\bae Z_\nu) W_\mu \big|^2  \nn\\
-\frac14 Z_\mn^2-\frac{\rho^2}{4} \big(g^2 W_\mu^*W_\mu
+\frac{g^2+g'^2}{2} Z_\mu^2 \big)   \nn\\
+i (e F_\mn^{\rm (em)}
+ \bae Z_\mn) W_\mu^* W_\nu  
+ \frac{g^2}{4}(W_\mu^* W_\nu - W_\nu^* W_\mu)^2,  \nn\\
D_\mu^{\rm (em)}=\pd_\mu+ieA_\mu^{\rm (em)},
\label{lag3}
\end{gather}
where $e$ and $\bae$ are the electric and neutral charges, 
\begin{gather}
e=\frac{gg'}{\sqrt{g^2+g'^2}},
~~~\bae =\frac{g^2}{\sqrt{g^2+g'^2}}.
\end{gather}
From this we have 
\begin{gather}
g= \sqrt{e^2+\bae^2},
~~~g' =	\frac{e}{\bae}~\sqrt{e^2+\bae^2}.
\end{gather}
Notice that $e/\bae=g'/g$.  Experimentally we have $\bae \simeq 1.83~e$, so that 
the weak coupling is stronger than the electromagnetic coupling. 

We emphasize that this is not the standard model in the unitary gauge. This is a gauge independent expression of the standard model. Moreover, this confirms that $W_\mu$ has both electric and neutral charge, so that it is doubly charged.

In the Lagrangian (\ref{lag3}) the Higgs doublet $\phi$ disappears completely. In stead we have the scalar 
Higgs particle which has the mass
\begin{gather}
M_H={\sqrt 2} \mu = \sqrt{\lam} \rho_0.	
\end{gather}
At the same time the W and Z gauge bosons acquire the mass
\begin{gather}
M_W= \frac{g}{2} \rho_0,   
~~~M_Z=\frac{\sqrt{(g^2+g'^2)}}{2}	\rho_0,
\end{gather}
absorbing the three physical degrees of the Higgs doublet. This, of course, is the Higgs mechanism. But here again 
we do not need any symmetry breaking 
to have the mass generation. Clearly (\ref{lag3}) is mathematically identical to (\ref{lag0}), so that it retains the full non-Abelian gauge symmetry of (\ref{lag0}). And here the W and Z bosons acquire mass from the non-vanishing vacuum value of 
the Higgs field $\rho_0$ which does 
not break any symmetry. This confirms that we do not need any spontaneous symmetry breaking to have the mass 
generation \cite{epjc15,pta19,epjc20}.  

From (\ref{lag3}) we obtain the following equations of motion
\begin{gather}
\pd^2 \rho-\big(\frac{e^2+\bae^2}{2} W_\mu^*W_\mu
+\frac{(e^2+\bae^2)^2}{4 \bae^2} Z_\mu^2 \big)~\rho  \nn\\
=\frac{\lam}{2}\big (\rho^2 -\rho_0^2 \big)~\rho,   \nn\\
\big(D_\mu^{\rm (em)}+i\bae Z_\mu\big) 
\big[(D_\mu^{\rm (em)}+i\bae Z_\mu) W_\nu \nn\\
-(D_\nu^{\rm (em)}+i\bae Z_\nu) W_\mu \big] 
=\frac{(e^2+\bae^2)}{4}\rho^2 W_\nu  \nn\\
+ i W_\mu \big(e F^{\rm(em)}_\mn 
+ \bae Z_\mn \big)  \nn\\
+ (e^2+\bae^2) W_\mu(W_\mu^* W_\nu-W_\nu^* W_\mu),\nn\\
\pd_\mu F_\mn^{\rm(em)} = J_\nu^{(e)},  \nn\\
\pd_\mu Z_\mn =J_\nu^{(n)},
\label{wseq2}
\end{gather}
where $J_\nu^{(e)}$ and $J_\nu^{(n)}$ are 
the electromagnetic and neutral currents 
given by
\begin{gather}
J_\nu^{(e)} = ie \Big\{\pd_\mu 
(W_\mu^* W_\nu-W_\nu^* W_\mu)  \nn\\
+W_\mu^* \big[(D_\mu^{\rm (em)} +i\bae Z_\mu) W_\nu 
-(D_\nu^{\rm (em)} +i\bae Z_\nu) W_\mu \big] \nn\\
-W_\mu \big[(D_\mu^{\rm (em)} +i\bae Z_\mu) W_\nu -(D_\nu^{\rm (em)} +i\bae Z_\nu) W_\mu \big]^*  \Big\},   \nn\\
J_\nu^{(n)} =i\bae \Big\{\pd_\mu  (W_\mu^* W_\nu 
-W_\mu W_\nu^*)  \nn\\
+ W_\mu^* \big[(D_\mu^{\rm (em)} +i\bae Z_\mu) W_\nu 
-(D_\nu^{\rm (em)} +i\bae Z_\nu) W_\mu \big] \nn\\
-W_\mu \big[(D_\mu^{\rm (em)} +i\bae Z_\mu) W_\nu -(D_\nu^{\rm (em)} 
+i\bae Z_\nu) W_\mu \big]^*  \Big\} \nn\\
+\frac{(e^2+\bae^2)^2}{4 \bae^2} \rho^2 Z_\nu.
\label{j2}
\end{gather}
This should be compared with the equations of motion (\ref{wseq1})  obtained from (\ref{lag2}). The two sets of currents in (\ref{j1}) and (\ref{j2}) are related by
\begin{gather}
\left(\begin{array}{cc} J_\mu^{\rm (e)} \\  J_\mu^{(n)} \end{array} \right)
=\frac{1}{\sqrt{g^2 + g'^2}} \left(\begin{array}{cc} g & g' \\
-g' & g \end{array} \right)
\left( \begin{array}{cc} j_\mu^{(h)} \\ 
j_\mu^{(w)} \end{array} \right).
\label{jmix}
\end{gather}
The existence of two conserved current in the standard model has been well known. But the fact that two sets of currents, the weak and hypercharge currents $j_\mu^{(w)}$ and $j_\mu^{(h)}$ and the electric 
and neutral currents $J_\mu^{(e)}$ and $J_\mu^{(n)}$, are related by the mixing (\ref{jmix}) has not been well appreciated.  

It has been asserted that the standard model has no monopole topology, and thus has no monopole \cite{col,vach,
bar}. However, the above exercise 
tells that this is not true. In fact the standard model has two independent monopole topologies. In the Abelianized Lagrangian (\ref{lag2}) both $A_\mu'$ and $B_\mu$ accommodate the Abelian monopole topology $\pi_1(S^1)$ separately. And these two Abelian monopole topologies translate to the two Abelian monopole topologies in $A_\mu^{\rm (em)}$ and $Z_\mu$.  

The existence of two different monopole topologies in the standard model stems from the fact that the SU(2) gauge potential $\A_\mu$ has the non-Abelian $\pi_2(S^2)$ monopole topology and the U(1) gauge potential $B_\mu$ has the Abelian $\pi_1(S^1)$ monopole topology. After the Abelian decomposition, the non-Abelian monopole topology changes to the Abelian monopole topology, so that the theory has two Abelian monopole topology. This strongly implies that the theory could accommodate two different types of monopoles. 

\section{A New Cho-Maison Monopole}

It is well known that the standard 
model has the Cho-Maison monopoles
which carry the magnetic charge $4\pi/e$, the naked one and the one dressed by W and Higgs bosons, \cite{chen,epjc15,ellis,bb,mav,pta19,
epjc20,gv,ms}. Now, we show that it has another Cho-Maison monopole which has only the W boson dressing. This tells that actually we have three Cho-Maison monopoles in the standard model.  

To show this we start from the Cho-Maison monopole ansatz in the spherical coordinates $(t,r,\theta,
\varphi)$ \cite{plb97,yang}
\begin{gather}
\phi=\frac{1}{\sqrt 2} \rho(r)~\xi,
~~~\xi =i \left(\begin{array}{cc}
\sin \dfrac{\theta}{2}~\exp(-i\varphi) \\
-\cos \dfrac{\theta}{2} \end{array} \right), \nn\\
\hA_\mu =-\frac1g~\hr \times \pd_\mu \hr, 
~~~\W_\mu =\frac{f(r)}g~\hr \times \pd_\mu \hr,    \nn\\
B_{\mu} =-\frac{1}{g'}(1-\cos\theta) 
\pd_\mu \vphi.
\label{cmmans0}
\end{gather}
Notice that $\A_\mu$ represents the 'tHooft-Polyakov monopole ansatz and $B_\mu$ describes the Dirac monopole. 
In terms of physical fields the ansatz becomes \cite{plb97,yang,pta19,epjc20}
\begin{gather}
\rho =\rho(r),  
~~~W_\mu= \frac{i}{g} \frac{f}{\sqrt 2} (\pd_\mu \theta +i\sin\theta \pd_\mu \vphi), \nn\\
A_\mu^{\rm (em)}
=-\frac1e (1-\cos\theta)~\pd_\mu \vphi,
~~~Z_\mu= 0.
\label{cmmans2}
\end{gather}
This clearly shows that the ansatz is for the electroweak monopole.

Now, we can use the equations of motion (\ref{wseq2}) to derive the equation for 
the Cho-Maison monopole. The ansatz (\ref{cmmans2}) automatically satisfies 
the last two equations of (\ref{wseq2}), 
and reduces the first two equations to 
the following equations,   
\begin{gather}
\ddot \rho+\frac{2}{r}\dot \rho -\frac{f^2}{2r^2} \rho
=\frac{\lam}{2} \big(\rho^2
-\rho_0^2 \big) \rho, \nn \\
\ddot f -\frac{f^2 -1}{r^2} f
=\frac{g^2}{4} \rho^2~f.
\label{cmeq}
\end{gather}
This has the naked Cho-Maison monopole solution
\begin{gather}
\rho=\rho_0 =\sqrt{2\mu^2/\lam},  
~~~A_\mu^{\rm (em)} =-\frac{1}{e}
(1-\cos \theta) \pd_\mu \varphi,  \nn\\
f=0,~~~Z_\mu=0,
\label{cmon}
\end{gather}
which has the magnetic charge $4\pi/e$ (not $2\pi/e$). This tells that the Cho-Maison monopole is a Schwinger (not Dirac) type monopole \cite{plb97}.

Moreover, with the boundary condition
\begin{gather}
\rho(0)=0,~~~~\rho(\infty)=\rho_0,  \nn\\
f(0)=1,~~~~f(\infty)=0,
\label{bc0}
\end{gather}
we can integrate (\ref{cmeq}) and have the electroweak monopole carrying the magnetic charge $q_m=4\pi/e$, with a non-trivial Higgs and W boson dressing. The monopole solution is shown in 
Fig. \ref{ncmm} in black curves \cite{plb97,epjc15}. 

Classically the Cho-Maison monopole 
has the following energy \cite{plb97,
epjc15,pta19},
\begin{gather}
E =E_0 +E_\rho +E_W +E_{\rho W},  \nn\\	
E_0 = \frac{4\pi}{{g'}^2} \Int_0^\infty ~\frac{dr}{2 r^2},  \nn\\
E_\rho =M \Int_0^\infty 2 r^2
\Big[\big(\frac{\dot \rho}{\rho_0} \big)^2 +\frac{\lam \rho_0^2}{4} \big((\frac{\rho}{\rho_0})^2 
-1 \big)^2 \Big] M_W dr, \nn\\
E_W= M \Int_0^\infty 
\Big[\dot f^2 +\frac{(f^2-1)^2}{2r^2} \Big] \frac{dr}{M_W}, \nn\\
E_{\rho W} = M \Int_0^\infty f^2 (\frac{\rho}{\rho_0})^2 M_W dr,   \nn\\
M =\frac{4\pi}{g^2}~M_W
\simeq 2.6~{\rm TeV}.
\label{cme}
\end{gather}
The boundary condition (\ref{bc0}) guarantees that $E_\rho$, $E_W$, and 
$E_{\rho W}$ are finite. But $E_0$ becomes infinite, so that it has infinite energy. This is because 
$B_\mu$ in (\ref{cmmans0}) has the monopole point singularity at the origin.  

There have been many efforts to estimate the monopole mass. Zeldovich and Khlopov first predicted the mass to be around 
11 TeV, $1/\al$ times the W boson mass, arguing that the monopole mass comes 
from the same Higgs mechanism which makes the W boson massive, except 
that here the coupling is 
$4\pi/e$ \cite{zel}. After this there have been other estimates, notably based on the scaling argument and the charge screening mechanism, and the general consensus now is that the mass could be around 4 to 11 TeV \cite{epjc15,pta19,ellis,bb,zel}. 

Now we can show that the standard model has a new Cho-Maison type monopole which has only the W boson dressing. To do
this, notice that in the absence of 
the Higgs doublet $\phi$ (i.e., when $\rho=0$) the Cho-Maison monopole equation (\ref{cmeq}) reduces to (\ref{qcdmeq}), the chromonic monopole equation in QCD. Obviously we can integrate this with the boundary condition (\ref{qcdmbc}), and obtain a new Cho-Maison monopole with magnetic charge $4\pi/e$ which has only the W boson dressing. This means that we have a new Cho-Maison monopole whose W boson profile is identical to the chromon profile of the new monopole in QCD. The monopole solution is shown in Fig. \ref{ncmm} in red curve.  

The energy of this monopole is given by
\begin{gather}
E =E_{CM} +E_W,  \nn\\	
E_{CM} =\frac{4\pi}{e^2} \Int_0^\infty \frac{dr}{2r^2},  \nn\\
E_W =\frac{4\pi}{g^2} \Int_0^\infty 
\Big[\dot f^2 + \frac{f^2-2}{2r^2} f^2  \Big] dr.
\label{cme1}
\end{gather}
Notice that, unlike (\ref{cme}) both $E_{CM}$ and $E_W$ become divergent here, because the boundary condition (\ref{qcdmbc}) makes $E_W$ divergent near the origin. Clearly this energy is different from the energy of the chromonic monopole (\ref{qcdme})
in QCD. This tells that, in spite of the fact that the W boson profile of this monopole is the same as that of the chromonic monopole, they have different energy. 

One might wonder if the naked Cho-Maison monopole can have only the Higgs dressing. To see this, notice that 
with $f=0$ the second equation of (\ref{cmeq}) is trivially satisfied. 
So we can try to find a solution which has only the Higgs dressing integrating the first equation of (\ref{cmeq}) with the boundary condition
\begin{gather}
\rho(0) = 0,~~~\rho(\infty) =\rho_0.
\end{gather}
But this does not lead to a meaningful solution, which means that such a solution does not exist.

\section{Neutral Monopole in the Standard Model}

So far the known electroweak monopoles have been the Cho-Maison monopoles, the naked one and the one dressed by the Higgs and W bosons. But the above result tells that we have a new Cho-Maison monopole which has only the W boson dressing. 

Now, we are ready to show that the standard model has a totally new type 
of electroweak monopoles, the electromagnetically neutral monopoles which carry the neutral magnetic charge $4\pi/\bae$. To do that we choose the ansatz
\begin{gather}
\phi =0,   \nn\\ 
\hA_\mu =-\frac1g~\hr \times \pd_\mu \hr, 
~~~\W_\mu =\frac{1}{g} f(r)~\hr \times \pd_\mu \hr,  \nn\\
B_\mu =\frac{g'}{g^2} (1-\cos \theta)
\pd_\mu \vphi.
\label{nans0}
\end{gather}
Notice that the ansatz for $\A_\mu$ is identical to the Cho-Maison monopole ansatz. But here we have no Higgs doublet, and $B_\mu$ has a different coefficient.

In terms of the physical fields, 
the ansatz is expressed by
\begin{gather}
\rho =0,
~~~W_\mu= \frac{i}{g} \frac{f}{\sqrt 2} (\pd_\mu \theta 
+i \sin\theta \pd_\mu \vphi),  \nn\\
A_\mu^{\rm (em)} =0,
~~~Z_\mu=-\frac{1}{\bae}(1-\cos \theta)
\pd_\mu \vphi.
\label{nans1}
\end{gather}
Notice that here again $\rho=0$ means no Higgs field. This should be compared with the Cho-Maison monopole ansatz (\ref{cmmans2}). The difference is that here $Z_\mu$, not $A_\mu^{(em)}$, contains the monopole which carries the neutral magnetic charge $4\pi/\bae$. This clearly tells that the ansatz is for the neutral monopole. 

\begin{figure}
\includegraphics[height=4.5cm, width=8cm]{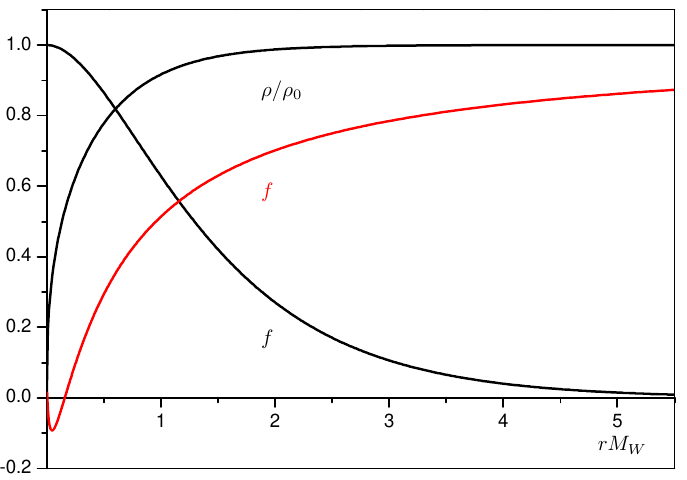}
\caption{\label{ncmm} The new monopole solutions in the standard model. The W boson profile of the neutral magnetic monopole is shown in red curve, which should be compared with the W and Higgs boson profiles of the Cho-Maison monopole shown in black curves. Notice that the same red curve also describes the W boson profile of the new Cho-Maison monopole which has only the W boson dressing.}
\end{figure}

With this we have the equation of motion from (\ref{wseq2}),
\begin{gather}
\ddot f -\frac{f^2 -1}{r^2} f =0.
\label{ncmeq}
\end{gather}
Obviously this is identical to the monopole equation (\ref{qcdmeq}) 
in QCD and the monopole equation (\ref{ntpmeq}) in Georgi-Glashow model. It has the trivial solution $f=0$, which describes the naked neutral monopole which carries the magnetic charge $4\pi/\bae$. Moreover, we can integrate the above equation with the boundary condition 
\begin{gather}	
f(0) = 0,~~~f(\infty) =1,
\label{nmbc}
\end{gather}
and obtain the neutral magnetic monopole solution which has the W boson dressing, which is identical to the chromonic monopole in QCD. The result is shown in 
Fig. \ref{ncmm} in red curves.
 
Notice that the boundary condition for the neutral monopole (\ref{nmbc}) is different from
the boundary condition we have in the Cho-Maison monopole,
\begin{gather}	
f(0) = 1,~~~f(\infty) =0.
\label{cmmbc}
\end{gather}
So, one might wonder if we can also have the neutral monopole with this boundary condition. This turns out to be impossible. Mathematically one can show that (\ref{ncmeq}) has no solution which satisfies the above boundary condition \cite{yang1}. 

\begin{figure}
\includegraphics[height=4.5cm, width=8cm]{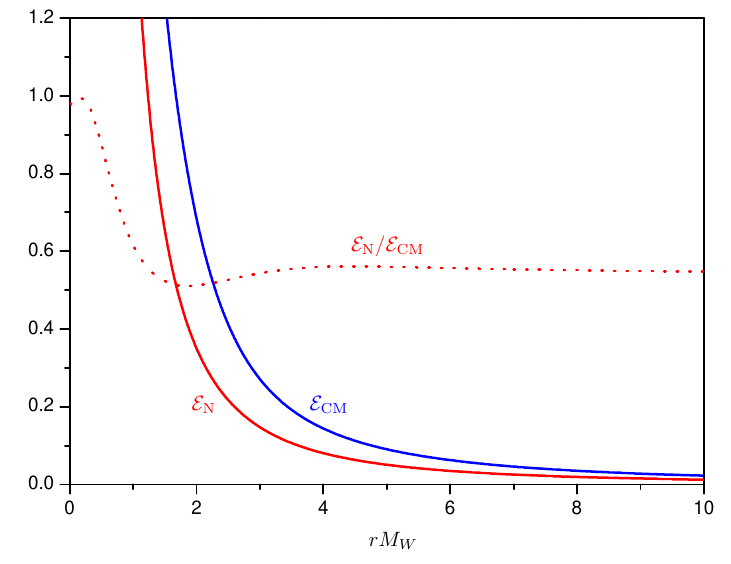}
\caption{\label{enden} The energy density of the electroweak monopoles. The blue curve represents the energy density of the Cho-Maison monopole which has the W and Higgs boson dressing, and the red curve represents the energy density of the neutral monopole which has the W boson dressing. The dotted line represents the ratio of the two energy densities.}
\end{figure}

We can easily calculate the magnetic charge of the neutral monopole. From
the ansatz (\ref{nans1}) we have
\begin{gather}
q_m = \int_{r=\infty}
Z_{\theta \vphi} d\theta d \vphi 
= -\frac{4\pi}{\bae},   \nn\\
Z_\mn= -\frac1{\bae} \sin \theta 
\big(\pd_\mu \theta \pd_\nu \vphi
-\pd_\nu \theta \pd_\mu \vphi \big),
\label{nmc}
\end{gather}
which tells that the monopole carries the neutral magnetic charge $-4\pi/\bae$.
Notice that, just like the Wu-Yang monopole dressed by the chromon in QCD, the W boson dressing here does not affect the magnetic charge of the monopole.

The neutral monopole has the energy 
\begin{gather}
E =E_N +E_W,  \nn\\	
E_N = \frac{4\pi}{\bae^2} \int_0^\infty \frac{dr}{2r^2},  \nn\\
E_W= \frac{4\pi}{g^2}\int_0^\infty 
\Big[\dot f^2 +\frac{f^2-2}{2r^2} f^2 \Big] dr,
\label{nme}
\end{gather}
where $E_N$ and $E_W$ are the energies of the singular neutral monopole and W boson dressing. Here again, both $E_N$ 
and $E_W$ become divergent near the origin. So, just like the Cho-Maison monopole, the neutral monopole also has an infinite energy classically. 

We can compare this energy with the energy of the Cho-Maison monopole
given by (\ref{cme}). This is shown in Fig. \ref{enden}, which clearly shows that the neutral monopole has less energy than the Cho-Maison monopole.   

How can we estimate the mass of the neutral monopole? There may be various ways to do that. For example, we could also assume the mass to be roughly $1/\bal$ times bigger than the W boson mass, where $\bal=\bae^2/4\pi$ is the fine structure constant of the neutral charge. With $\bae \simeq 1.83 \times e$, we have $\bal \simeq 3.35 \times \al$. This implies that the neutral monopole mass could be around 3.23 TeV, much smaller than the mass of the Cho-Maison monopole. Of course we need a more reliable estimate. But the energy density shown in Fig. \ref{enden} clearly shows that the mass is less than that of the Cho-Maison monopole.  

To remove any lingering doubt on the existence of the new monopole, we now show why such monopole should exist in the standard model. To do this we have to remember how Dirac constructed the 
Dirac monopole in Maxwell's theory.
It is well known that Dirac monopole originates from the Dirac string, the infinitely thin string carrying the quantized magnetic flux $2\pi/e$. When we cut it, the monopole flux appears at the cutted edge. But this is not a monopole but a magnetic flux coming out of the string. However, Dirac showed us how to make this string invisible with the Dirac's charge quantization condition, and change this to the Dirac monopole.

\begin{figure}
	\includegraphics[width=0.3\linewidth]{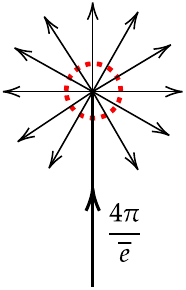}
	\caption{\label{namo} The infinitely thin Z string carrying the neutral magnetic flux $4\pi/\bar e$ along the negative $z$-axis. Making the neutral 
		string invisible with the neutral charge quantization condition, we obtain the neutral monopole in the standard model shown in dotted red circle.}
\end{figure}

If so, we can also construct the neutral monopole from the neutral string which carries the neutral magnetic flux $4\pi/\bae$, imposing the neutral charge quantization condition. This is shown in Fig. \ref{namo}. But do we have 
the neutral magnetic flux string in the standard model? Yes, we do. To show this, notice that (\ref{wseq2}) has 
the solution
\begin{gather}
\rho=0,~~~A_\mu^{\rm (em)}=0, 
~~~W_\mu=0,   \nn\\
Z_\mu=\frac{n}{\bae} \pd_\mu \varphi,
\label{zs}	
\end{gather}	
where $n$ is an integer. This is the singular Z string in the standard model
which carries the neutral magnetic flux $4\pi/\bae$.

So, with the neutral charge quantization condition $\bae =4\pi n/\bar g$, we have the naked neutral monopole shown by the dotted circle in Fig. \ref{namo}. This tells that the standard model must have the neutral monopole. Moreover, it must have two charge quantization conditions, the neutral charge quantization and the Dirac charge quantization conditions.

Interestingly, Nambu told us that the standard model has the Z string, a neutral string carrying the quantized neutral magnetic flux \cite{nambu,vacha,
epjc24}. Cutting this neutral string, 
we have the neutral monopole flux coming out of the neutral string. This is called the ``Nambu monopole" in the literature \cite{kib,eto}. But the Nambu's Z string is not a singular string. Moreover, the Nambu's monopole is not a monopole, but a neutral magnetic flux coming out of the string.

Before we leave this section, we show that the two monopoles, the Cho-Maison monopole which has the W boson dressing and the neutral monopole which has the same W boson dressing, come from another set of  monopoles in the standard model, the weak monopole of the SU(2) weak interaction and the hypercharge monopole of the hypercharge U(1) interaction. To show this, we have to show the existence of the weak monopole and the hypercharge 
monopole in the standard model first. 

Let us choose the following ansatz
for the SU(2) monopole,
\begin{gather}
\phi=0,~~~\hA_\mu = -\frac1g \hr \times \pd_\mu \hr,   \nn\\
\W_\mu =\frac{f(r)}{g} ~\hr \times \pd_\mu \hr,~~~B_\mu =0,
\label{wmans0}
\end{gather}
which can be expressed by
\begin{gather}
\rho=0,~~~A_\mu'
=-\frac1g (1-\cos \theta)~\pd_\mu \vphi, \nn\\
W_\mu= \frac{i}{g} \frac{f}{\sqrt 2} (\pd_\mu \theta +i \sin\theta \pd_\mu \vphi), 	
~~~B_\mu =0. 
\label{wmans1}
\end{gather}
This shows that the ansatz is for the SU(2) weak monopole.

From the ansatz, we have the following equation of motion
\begin{gather}
\ddot{f}-\frac{f^2-1}{r^2} f=0,
\label{wmeq}
\end{gather}
which is identical to (\ref{ncmeq}).
Integrating this with the boundary condition (\ref{nmbc}), we have the weak monopole which has the same W boson dressing as the neutral monopole with the W boson dressing. This confirms that the standard model has the weak monopole which has the W boson dressing. 

Now we discuss the hypercharge monopole. 
For this we choose the ansatz
\begin{gather}
\phi=0,~~~\hA_\mu =0,~~~\W_\mu =0,  \nn\\
B_\mu =-\frac1{g'} (1-\cos \theta) \pd_\mu \vphi.
\label{hmans0}
\end{gather}
Obviously this is the ansatz for the hypercharge monopole. We can easily show that this is another solution, the hypercharge monopole solution, of the standard model.

One might wonder if these monopole solutions are another set of monopole
solutions in the standard model. But they are not the new monopole solutions but a different realization of the two monopoles we discussed before, the Cho-Maison monopole which has the W boson dressing and the neutral monopole which has the same W boson dressing. To see this, notice that the ansatz $A_\mu^{\rm (em)}$ and $Z_\mu$ in (\ref{cmmans2}) and (\ref{nans1}) are the linear combinations of $A_\mu'$ and $B_\mu$ in the ansatz (\ref{wmans1}) and (\ref{hmans0}). Moreover, the currents of two potentials are related by the mixing (\ref{jmix}), so that the W boson dressing comes from the same source, the W boson ansatz in (\ref{wmans1}). This is why the two monopoles have the same W boson dressing.

This tells that we can indeed interpret 
the weak monopole and the hypercharge monopole as a different realization of the Cho-Maison monopole with the W boson dressing and the neutral monopole with the same W boson dressing. Clearly this interpretation would have been impossible without the existence of the mixing (\ref{jmix}) between the currents of two potentials $A_\mu'$ and $B_\mu$. 

\section{Discussion}

In this paper we have discussed the new monopole solutions in QCD, the Georgi-Glashow model, and the standard model. In QCD we have shown that there 
is a new monopole, the Wu-Yang monopole which has the chromon dressing, in sddition to the well known Wu-Yang monopole.

In the Georgi-Glashow model, we have shown that there are three monopoles, the naked Wu-Yang monopole, the 'tHooft-Polyakov monopole, and the Wu-Yang monopole which has only the W boson dressing. The first two monopoles are well known, but the last one is the new monopole. In doing so we have also shown that the 'tHooft-Polyakov monopole is nothing but the Wu-Yang monopole which has the Higgs and W boson dressing, which makes the energy finite. 

Moreover, in the standard model we have shown that there are actually two types electroweak monopoles, the Cho-Maison type which has the magnetic charge $4\pi/e$ and the neutral type which has the neutral magnetic charge $4\pi/\bae$. In specific we have shown that there are three different Cho-Maison monopoles, the naked one, the one which has the Higgs and W boson dressing, and the one which has only the W boson dressing. Moreover, we have shown that there are two neutral monopoles, the naked one and the one which has the W boson dressing. So far, we have had only only one type of electroweak monopole in the standard model, the naked Cho-Maison monopole and the one dressed by W and Higgs Bosons. But now we have three more new monopoles in the standard model, the Cho-Maison monopole which has only the W boson dressing, the naked neutral monopole, and the neutral monopole which has only the W boson dressing.

The existence of two types of monopoles in the standard model is completely unexpected. But this stems from the fact that the standard model has two conserved charges, the electric charge $e$ and the neutral charge $\bae$. This, of course, does not guarantee the existence of two types of monopoles in the standard model. So the fact that the theory actually accommodates two types of monopoles is really remarkable.  

We have also shown that the two types of monopoles in the standard model can also be represented by two sets of new monopoles, the weak monopole of SU(2) and the hypercharge monopole of U(1). 
And the Weinberg mixing transforms these monopoles to the Cho-Maison monopole and the neutral monopole.

All new monopoles have the common ancestor, the Wu-Yang monopole dressed by the chromon in QCD. Indeed all the new momopoles are the embedding of the chromonic monopole 
to different non-Abelian gauge theories. This tells that the chromonic monopole in QCD is the prototype of the new monopoles, although physically they are different objects.   

From the theoretical point of view the new monopoles are all very interesting, but the most important ones are the monopoles in the standard model. This is because they are the most realistic monopole which could exist in nature and thus can actually be detected. In particular, the neutral magnetic monopoles are a totally new type of monopoles whose existence has never been suggested in physics before.  
 
The standard model has really remarkable features. Obviously it is one of the most successful theories in high energy physics. Strangely, however, it has been asserted that the standard model has no monopole topology, and thus has no monopole \cite{col,vach,bar}. This is not true. 

Our analysis in this paper tells that the standard model has not just one but actually two monopole topologies, the SU(2) monopole topology $\pi_2(S^2)$ and the hypercharge U(1) monopole topology $\pi_1(S^1)$, which transform to two $\pi_1(S^1)$ monopole topology after the Abelian decomposition, the $\pi_1(S^1)$ monopole topology of the weak U(1) subgroup of SU(2) and the $\pi_1(S^1)$ monopole topology of the hypercharge U(1). This  justifies the existence of two types of monopoles in the standard model. This is completely unexpected.

One might ask if these new monopoles 
are stable. Obviously they have the topological stability. But one might like to know if they are also dynamically stable. It has been shown that the Cho-Maison monopole is dynamically stable \cite{gv}. It would be interesting to see if one can also prove the dynamical stability of the neutral monopole in the standard model.

The importance of the standard model may not be limited to high energy physics. It could also play important roles in low energy physics. This is because the standard model could be viewed as the non-Abelian Ginzburg-Landau theory of two-gap ferromagnetic superconductivity \cite{pla23,ap24,arx251}. Moreover, 
the standard model could also be viewed to describe the magnon electron spintronics \cite{arx251,arx252,arx253}.
This is because the Higgs doublet could be interpreted as the charged spinon which describes the electron in electron spintronics. This strongly implies that the two types of electroweak monopoles could also play important roles in condensed matter physics.   

The existence of two types of monopoles strongly implies that the standard model has new territories which have yet to be explored. The details of the new monopoles and related topological objects will be discussed in a separate paper \cite{cho}.

{\bf Acknowledgments}

~~~We sincerely thank Yisong Yang for sharing the existence proof of the neutral magnetic monopole with us. LZ and PZ are supported by the National 
Key R\&D Program of China (No. 2024YFE0109802) and National Natural Science Foundation of China (Grant No. 12175320 and No. 12375084). YMC is supported in part by the President's Fellowship Initiative of Chinese Academy of Sciences (Grant No. 2025PD0115).

\end{document}